\RequirePackage{fix-cm}
\documentclass[smallextended]{svjour3}       
\smartqed  
\usepackage{graphicx}
\usepackage{amsmath}
\usepackage[hidelinks]{hyperref} 
\journalname{Computing and Software for Big Science}
\begin{document}

\title{A pattern recognition algorithm for quantum annealers
}


\author{Fr\'{e}d\'{e}ric Bapst \and Wahid Bhimji  \and Paolo Calafiura \and Heather Gray  \and Wim Lavrijsen  \and Lucy Linder\thanks{Corresponding author. \email{lucy.linder@hefr.ch}}}

\institute{
Fr\'{e}d\'{e}ric Bapst \and Lucy Linder \at
{\bf   School of Engineering and Architecture of Fribourg, Switzerland, HES-SO, iCoSys Institute}
\and
Wahid Bhimji \and Paolo Calafiura \and Heather Gray \and Wim Lavrijsen \at
{\bf Lawrence Berkeley National Laboratory\\
Berkeley, CA 94720, USA}
\and Heather Gray \at {\bf Department of Physics, University of California, Berkeley, CA 94720, USA} }



\date{Received: date / Accepted: date}

\maketitle

\begin{abstract}

 The reconstruction of charged particles will be a key computing challenge for the high-luminosity Large Hadron Collider (HL-LHC) where increased data rates lead to large increases in running time for current pattern recognition algorithms. An alternative approach explored here expresses pattern recognition as a Quadratic Unconstrained Binary Optimization (QUBO) using software and quantum annealing. At track densities comparable with current LHC conditions, our approach achieves physics performance competitive with state-of-the-art pattern recognition algorithms. More research will be needed to achieve comparable performance in HL-LHC conditions, as increasing track density decreases the purity of the QUBO track segment classifier.

\end{abstract}

\section{Introduction}
\label{intro}

Early quantum computers are rapidly being made available both in the cloud and as prototypes in academic and industrial settings. These devices span the range from D-Wave~\cite{DWAVE} commercial quantum annealers to gate-based quantum processor prototypes based on a wide range of promising technologies~\cite{Preskill18}. Quantum computing holds the potential for super-polynomial speedups and large decreases in energy usage, if suitable algorithms can be developed. It is therefore crucial to start identifying algorithms and applications for high energy physics, to be ready for when quantum computing becomes mainstream and to provide input about what features are needed in quantum computers to solve problems in high-energy physics.

The reconstruction of charged particles will be a key computing challenge for the high-luminosity Large Hadron Collider (HL-LHC) where increased data rates lead to large increases in running times for conventional pattern recognition algorithms. Conventional algorithms~\cite{Cornelissen:1020106,Cornelissen_2008,Speer:884424,Cucciarelli:2006mt}, which are based on combinatorial track seeding and building, scale quadratically or worse as a function of the detector occupancy.

We present an alternative approach, one that expresses pattern recognition as a Quadratic Unconstrained Binary Optimization (QUBO; a NP-hard problem) using annealing, a process to find the global minimum of an objective function -- in our case a quadratic function over binary variables. The term annealing is inspired by the metallurgic process of repeated heating and cooling to remove dislocations in the lattice structure.
Likewise as used here, the annealing optimization process uses random ``thermal" fluctuations to find better results of the objective function, combined with a ``cooling" which progressively reduces the probability of accepting a worse result. Quantum annealing is grounded in the adiabatic theorem: a system will remain in its eigenstate if perturbations that act on it are slow, and small enough not to span the gap between the ground and first excited states.
Thus, it is possible to initialize a quantum annealer with a simple ground state Hamiltonian and evolve it adiabatically to the desired, complex, problem Hamiltonian. After evolution, quantum fluctuations, such as tunneling, bring the annealer into the ground state of the latter, representing the global minimum solution of the problem. All steps of quantum annealing operate on the system as a whole and the total time required is therefore in principle independent of the size of the system. Thus, as long as the problem fits on the annealer, the total running time should be constant, and a large enough quantum system, running a large problem, should outperform a software-based one.

We test our approach using annealing both in software simulation and by running on a D-Wave quantum computer. We use a dataset representative of the expected conditions at the HL-LHC from the TrackML challenge~\cite{trackml}. We study the performance of the algorithm as a function of the particle multiplicity. We do not expect to obtain speed improvements because the size of the currently available annealers is smaller than the scale of our problem.

\section{Methodology}
\label{sec:3}

\subsection{Pattern recognition, general considerations}\label{methodology-general}

The goal of pattern recognition is to identify groups of detector \emph{hits} to form tracks. Track trajectories are parameterized using the following five parameters: $d_0, z_0, \phi_0, \cot{\theta},$ and $q/p_T$\footnote{We use a right-handed coordinate system with its origin at the nominal interaction point (IP) in the center of the detector. The $x$-axis points from the IP to the center of the LHC ring, the $y$-axis points upward, and the $z$-axis coincides with the
axis of the beam pipe.  Cylindrical coordinates ($r$,$\phi$) are used in the transverse plane, $\phi$ being the azimuthal angle around the beam pipe. The polar angle $\theta$ lies in the $r-z$ plane.}. The transverse impact parameter, $d_0$, is the distance of closest approach of the helix to the reference point in the $x-y$ plane. The longitudinal impact parameter, $z_0$, is the $z$ coordinate of the track at the point of closest approach. The azimuthal angle, $\phi_0$, is the angle between the track and the tangent at the point of closest approach. The polar angle, $\cot{\theta}$ is the inverse slope of the track in the $r-z$ plane.  The curvature, $q/p_T$, is the inverse of the transverse momentum with the sign determined by the charge of the particle.

Neglecting noise and multiple scattering, most particle tracks of physics interest, particularly those with high $p_T$, exhibit the following properties:

\begin{itemize}
\item
  the hits follow an arc of a helix in the $x-y$ plane with a large radius of curvature or small $q/p_T$;
\item
  the hits follow a straight line in the $r-z$ plane
\item
  most hits lie on consecutive layers: there are few to no missing hits (holes)
\end{itemize}

Track candidates with fewer than five hits are predominantly fake tracks, which do not correspond to a true particle trajectory. While tracks can share hits, we impose the constraint from Ref.~\cite{trackml} that any one hit can belong to at most one track.

\subsection{Algorithm goals}\label{methodology-goals}

The algorithm presented in this paper encodes a classification problem. Following Ref.~\cite{stimpfl91}, tracks are constructed from $n-1$ doublets. Given the large set of potential doublets from hits in the detector, the goal is to determine which subset belongs to the trajectories of charged particles. The algorithm aims to preserve the efficiency, but improve the purity of the input doublet set.

\subsection{Triplets and quadruplets}\label{methodology-triplets}
We follow a similar approach to Ref.~\cite{stimpfl91}, but use triplets instead of doublets. In addition to improving the performance at high multiplicity, this allows us to calculate and use track properties.

A triplet, denoted $T^{abc}$, is a set of three hits ($a, b, c$) or a pair of consecutive doublets ($a, b$ and $b, c$), ordered by increasing transverse radius ($R$). Two triplets $T^{abc}$ (of hits $a, b, c$) and $T^{def}$ (of hits $d, e, f$), can be combined to form a quadruplet if $b=d \land c=e$ or a quintet if $c=d$. If they share any other hit, the triplets are marked as being in conflict. A set of $n$ consecutive hits will result in $n-2$ triplets and $n-3$ quadruplets.

\begin{figure}
\centering
\includegraphics[width=\textwidth]{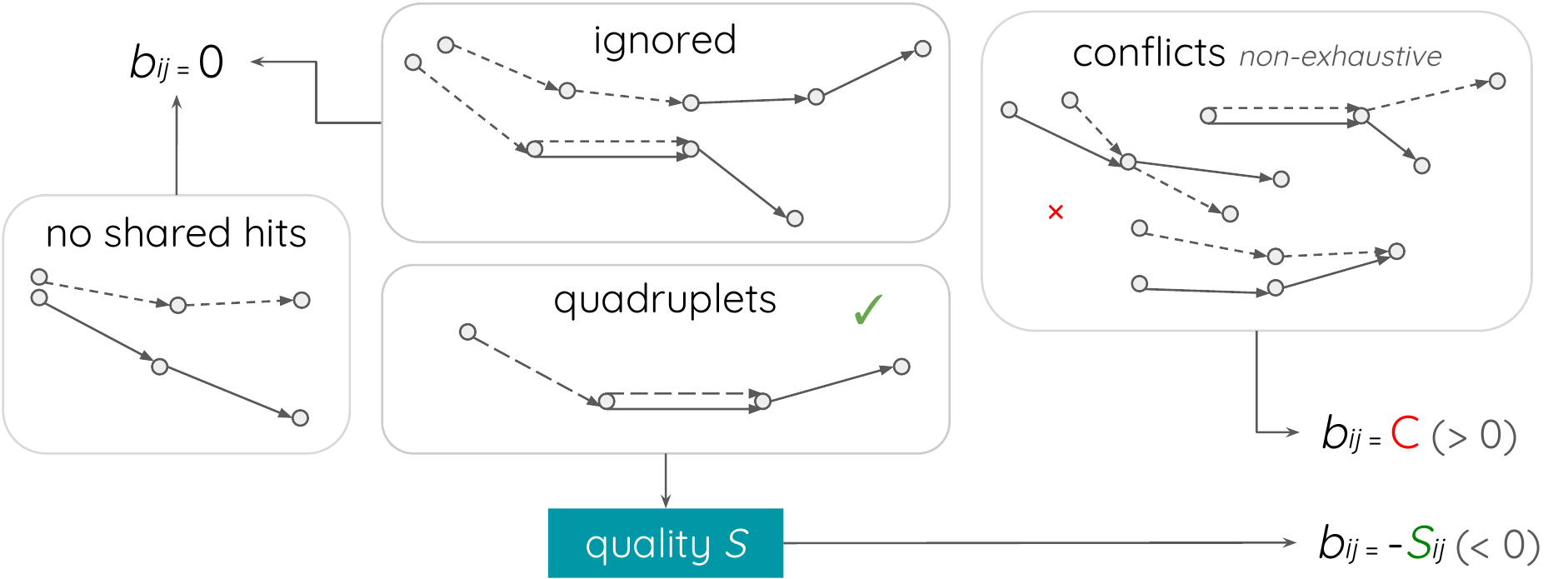}
\caption{The value assigned to the QUBO quadratic weights $b_{ij}$ for different configurations of the pairs of triplets $T_i$ and $T_j$. See text for details.}
\label{fig:plots-examples}
\end{figure}

Key triplet $T_{i}^{abc}$ properties are the number of holes $H_i$; the curvature, $
q/p_T$; and the angle $\delta\theta$ between the doublets.

The strength $S$ quantifies the compatibility of the track parameters between the two triplets in a quadruplet $(T_i, T_j)$:
\begin{equation}\label{eq:strength-generic}
S(T_i, T_j) =
{z_1} \frac{ {z_2} \big(1-|\delta(q/p_{Ti}, q/p_{Tj})|\big)^{z_3} + (1-z_2) \big(1-\mathrm{max}(\delta\theta_i, \delta\theta_j)\big)^{z_4}
}{
(1+H_i+H_j)^{z_5}
}
\end{equation}
where $z_2$ encodes the relative importance of the curvature with respect to $\delta\theta$. The other parameters ($z_1, z_3, z_4, z_5$) are unbounded constants that require problem-specific tuning. The parameters are set to favour high $p_T$ tracks. In its simplest form, we have $z_2 = 0.5$ (equal weights), $z_5 = 2$, and all other constants set to 1:
\begin{align}\label{eq:strength}
S(T_i, T_j)
&= \frac{1 - \frac{1}{2}(|\delta(q/p_{Ti}, q/p_{Tj})| + \mathrm{max}(\delta\theta_i, \delta\theta_j))}{(1+H_i+H_j)^2}
\end{align}

\subsubsection{Definition of the Quadratic Unconstrained Binary Optimization}\label{methodology-qubo}

The QUBO is configured to identify the best pairs of triplets.
It has two components: a linear term that weighs the quality of individual triplets and a quadratic term used to express relationships between pairs of triplets.
In our case, the objective function to minimize becomes:
\begin{equation}
O(a,b,T) = \sum_{i=1}^N{a_i T_i} + \sum_{i}^N\sum_{j < i}^N{b_{ij} T_i T_j}
\end{equation}
where $T$ are all potential \emph{triplets}, $a_i$ are the \emph{bias weights}, and $b_{ij}$ the \emph{coupling strengths} computed from the relation between the triplets $T_i$ and $T_j$. The bias weights and the coupling strengths define the Hamiltonian. Minimizing the QUBO is equivalent to finding the ground state of the Hamiltonian.

All bias weights are set to be identical $a_i=\alpha$ , which means all triplets have equal {\em a priori} probability to belong to a particle track. Our objective function therefore depends solely\footnote{No difference was observed when shifting the bias weight $\alpha$ by a small amount.} on the triplet-triplet coupling strength $b_{ij}$. If the triplets form a valid quadruplet, the coupling strength is negative and equal to the quadruplet quality $S(T_i, T_j)$ (Eq. \ref{eq:strength}). If the two triplets are in conflict, the coupling is a positive constant $b_{ij}=\zeta$ that disfavours a solution with $T_i=T_j=1$. Finally, if the triplets have no relationship (meaning,  no shared hits), the coupling is set to zero.
This is illustrated in Figure~\ref{fig:plots-examples} and represented in Equation~\ref{eq:coupling-strength}.
\begin{equation}\label{eq:coupling-strength}
b_{ij} =
\begin{cases}
	-S(Ti, Tj), 	& \mbox{if } (T_i, T_j) \mbox{ form a quadruplet}, \\
	\zeta 				& \mbox{if } (T_i, T_j) \mbox{ are in conflict}, \\
	0 					& \mbox{otherwise.}
\end{cases}
\end{equation}
As is clear from Eq.~\ref{eq:coupling-strength}, the choice of constants in Eq.~\ref{eq:strength-generic} determines the functional behavior of $b_{ij}$.
The larger the conflict strength $\zeta$ the lower the number of conflicts, but too large values risk discontinuities in the energy landscape, increasing time to convergence.
Furthermore, the D-Wave machines limit the value of $\zeta$ to between $-2$ and $2$ (with a restricted precision).

\subsection{Dataset Selection}\label{methodology-cuts}

By design the algorithm does not favor any particular momentum range.
However, to limit the size of the QUBO, we focus on high $p_T$ tracks ($ p_T >= 1 GeV$), which are most relevant for physics analysis at the HL-LHC.

A triplet $T_i$ is created if and only if:
\begin{eqnarray*}
  H_i & \leq & 1 \\
  |(q/p_T)_i | &\leq & 8*10^{-4} \and \\
  \delta\theta_i & \leq & 0.1
\end{eqnarray*}
And a quadruplet $(T_i, T_j)$ is created if and only if:
\begin{eqnarray*}
  |\delta ((q/p_T)_i, (q/p_T)_j)| & \leq & 1*10^{-4} \and \\
  S(T_i, T_j) & > & 0.2
\end{eqnarray*}

Triplets that are not part of any quadruplet or whose longest potential track has less than five hits, are not considered.

\section{Experimental setup}\label{sec:experimental-setup}

\subsection{Dataset}\label{sec:setup-dataset}

The TrackML dataset is representative of future high-energy physics experiments at the HL-LHC. It anticipates the HL-LHC multiplicities planned for after 2026. Both the low $p_T$ cut (150 MeV) and high luminosity (200 $\mu$) make pattern recognition within this dataset a challenging task. We simplify the dataset by focusing on the barrel region of the detector, i.e. hits in the end-caps are removed. If a particle makes multiple energy deposits in a single layer, all but one energy deposits are removed. Hits from particles with $p_T < 1$~GeV and particles with less than five hits are kept and thus part of the pattern recognition, but are not taken into account when computing the performance metrics.
Events are split by randomly selecting a fraction of particles and an equal fraction of noise to generate datasets with different detector occupancies yet similar characteristics.

\subsection{Metrics}\label{sec:setup-metrics}

The performance is assessed using {\em purity} and {\em efficiency}\footnote{Instead of purity and efficiency, the equivalent terms of precision and recall are sometimes used in the literature.}, which are computed on the final set of doublets. This provides a good estimate of the quality of the model as a doublet classifier, but does not
account for the difference in importance between track candidates for physics. The TrackML score~\cite{trackml} is used as a complementary metric as it includes weights to favour tracks with higher $p_T$, which play a larger role in physics performance.

The efficiency and purity are defined as follows:

\begin{equation}
    \mathrm{efficiency} = \frac{D^{\mathrm{rec}}_{\mathrm{matched}}}{D^{\mathrm{true}}}
\end{equation}

\begin{equation}
    \mathrm{purity} = \frac{D^{\mathrm{rec}}_{\mathrm{matched}}}{D^{\mathrm{rec}} - D^{\mathrm{rec}}_{\mathrm{oa}}}
\end{equation}
The number of true doublets ($D^{true}$) only includes those with $p_T > 1$~GeV, which deposit at least five hits in the detector barrel. Reconstructed doublets ($D^{rec}$) are matched to true doublets using truth information ($D^{rec}_{\mathrm{matched}}$. Reconstructed doublets matched to true doublets, but with either $p_T~\leq~1$~GeV or less than five hits in the detector barrel ($D^{rec}_{oa}$) are excluded from the purity.

\subsection{Initial doublets}\label{sec:initial-doublets}

The initial set of doublets is generated using an adaptation as a Python library of the ATLAS online track seeding code~\cite{delgado2016atlas}.
It was tuned to ensure an efficiency above 99\% for high $p_T$ tracks, but has a purity below 0.5\%.

\subsection{QUBO solver}\label{sec:setup-solver}

qbsolv~\cite{qbsolv} is a tool developed by D-Wave to solve larger and more densely connected QUBOs than currently supported by the D-Wave hardware. It uses an iterative hybrid classical/quantum approach with multiple trials. In each trial, the QUBO is split into smaller instances that are submitted to a sub-QUBO solver for global optimization. Results are combined and a tabu search~\cite{Glover86} is performed for local optimization. The sub-QUBO solver is either a D-Wave system or a software-based solver.
Using this setup, running qbsolv on a classical system has the same workflow as running qbsolv with D-Wave, making it an effective simulator.
D-Wave also provides NEAL~\cite{NEAL}, a standalone software-only annealer, which we use for comparison studies.

The number of sub-QUBOs that are created can be controlled by restricting the size of the number of logical qubits that can be used per sub-QUBO.
We use the default value of $47$ for both the simulator and the D-Wave, as it worked well: larger or smaller numbers can result in a failed mapping, and a subsequent abort of the run.

We ran our simulations on the Cori~\cite{cori} supercomputer at NERSC, experiments on the Ising D-Wave 2X machine at Los Alamos National Laboratory (with $1000$ qubits), and tests on the D-Wave LEAP cloud service. The number of iterations and D-Wave samplings was limited to $10$.

\subsection{Complete algorithm}\label{methodology-pseudocode}

\begin{figure}
\centering
\includegraphics[width=\textwidth]{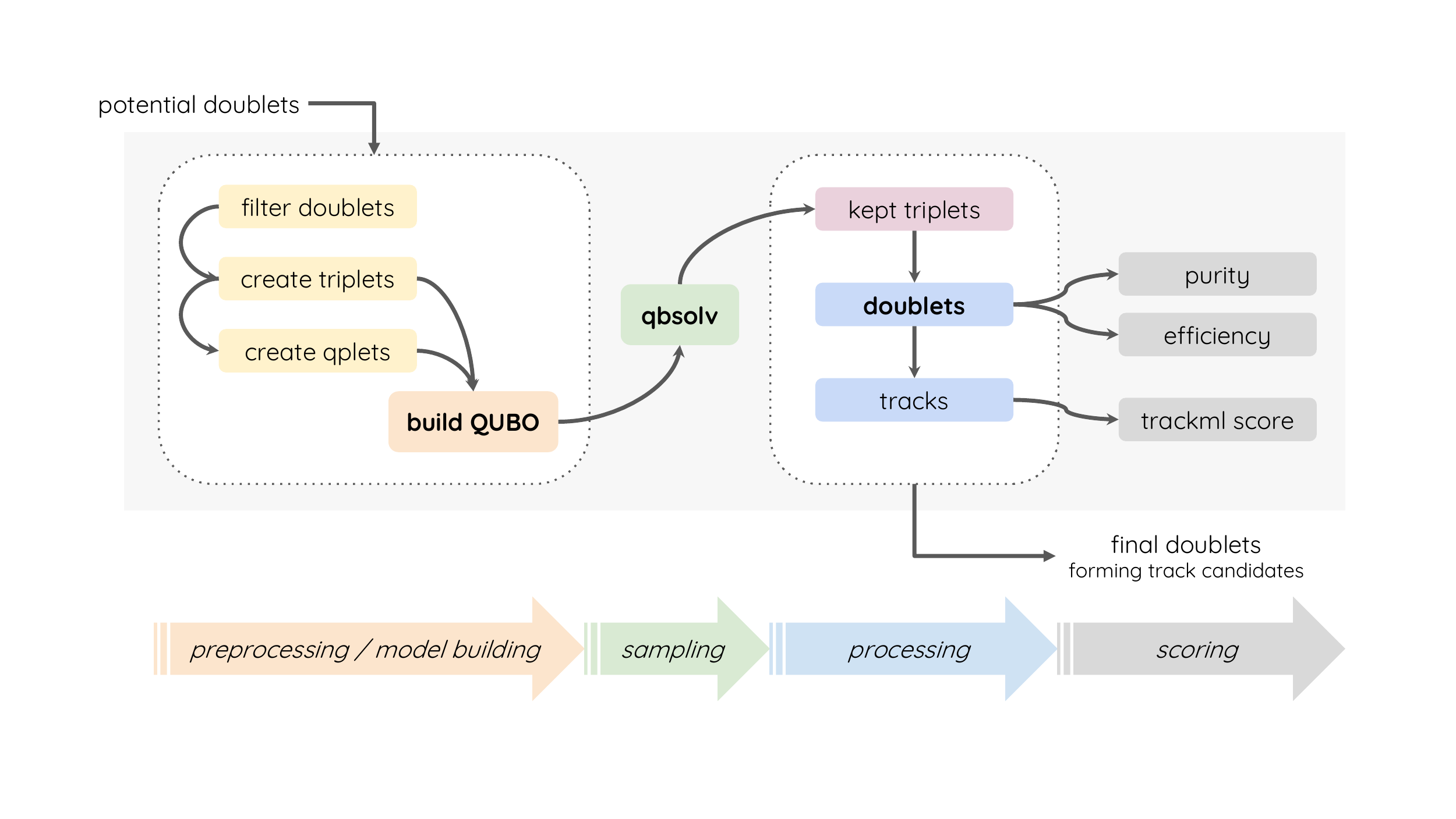}
\caption{Overview of the steps in the algorithm from pre-processing, sampling on the D-Wave, post-processsing and scoring.}
\label{fig:algo}
\end{figure}


Figure \ref{fig:algo} illustrates the steps in the algorithm.
The initial doublets are combined into triplets and quadruplets, after satisfying the requirements from Section~\ref{methodology-cuts}. The QUBO is generated and sampled using qbsolv. The post-processing phase includes converting the triplets into doublets, removing duplicates and dealing with any remaining conflicts. The track candidates are reconstructed from the doublets and track candidates with less than five hits are discarded. Finally, performance metrics are computed and the set of final doublets corresponding to the track candidates is returned.

\section{Results}
\label{sec:5}

We chose three events from the dataset containing 10K, 12K and 14K particles plus noise, with the latter being the highest multiplicity event in the dataset. We sample from these events to construct sets ranging from O(1K) to O(7K) particles. Each set is constructed by taking a fixed fraction of the particles and the noise in that event.

\subsection{Algorithmic Performance}
\label{sec:results-physics-performance}

\begin{figure}[ht!]
\centering
\includegraphics[width=\textwidth]{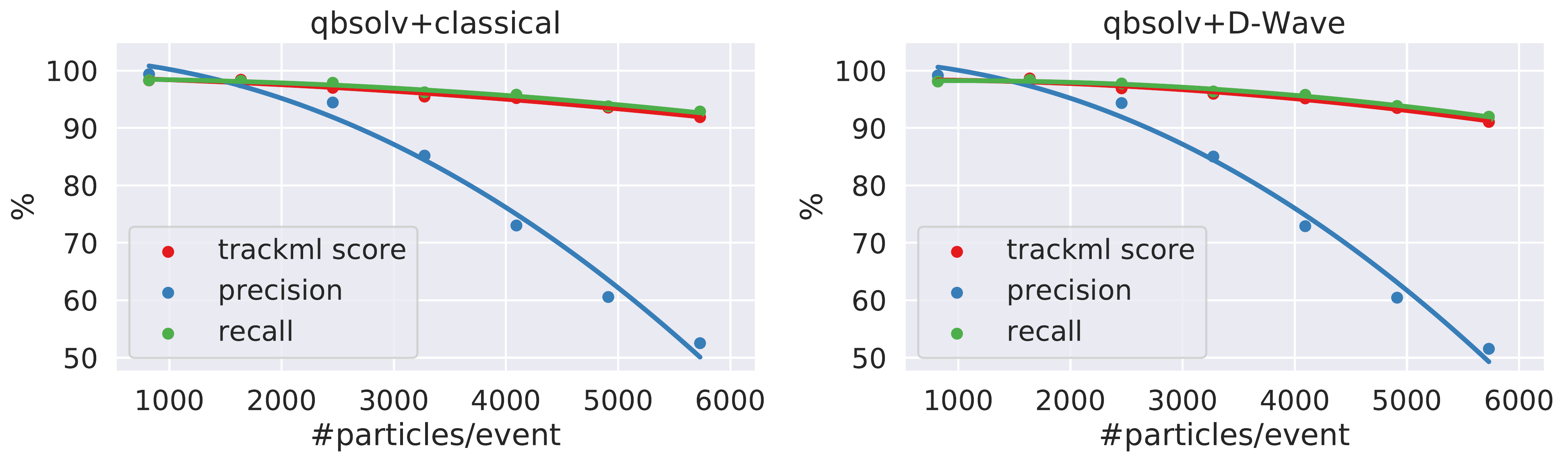}
\caption{The performance of classical simulator (left) and D-Wave (right), as measured by TrackML score (red), purity (blue), and efficiency (green), as a function of particle multiplicity.}
\label{fig:physics_performance}
\end{figure}

\begin{figure}[ht!]
\centering
\includegraphics[width=.55\textwidth]{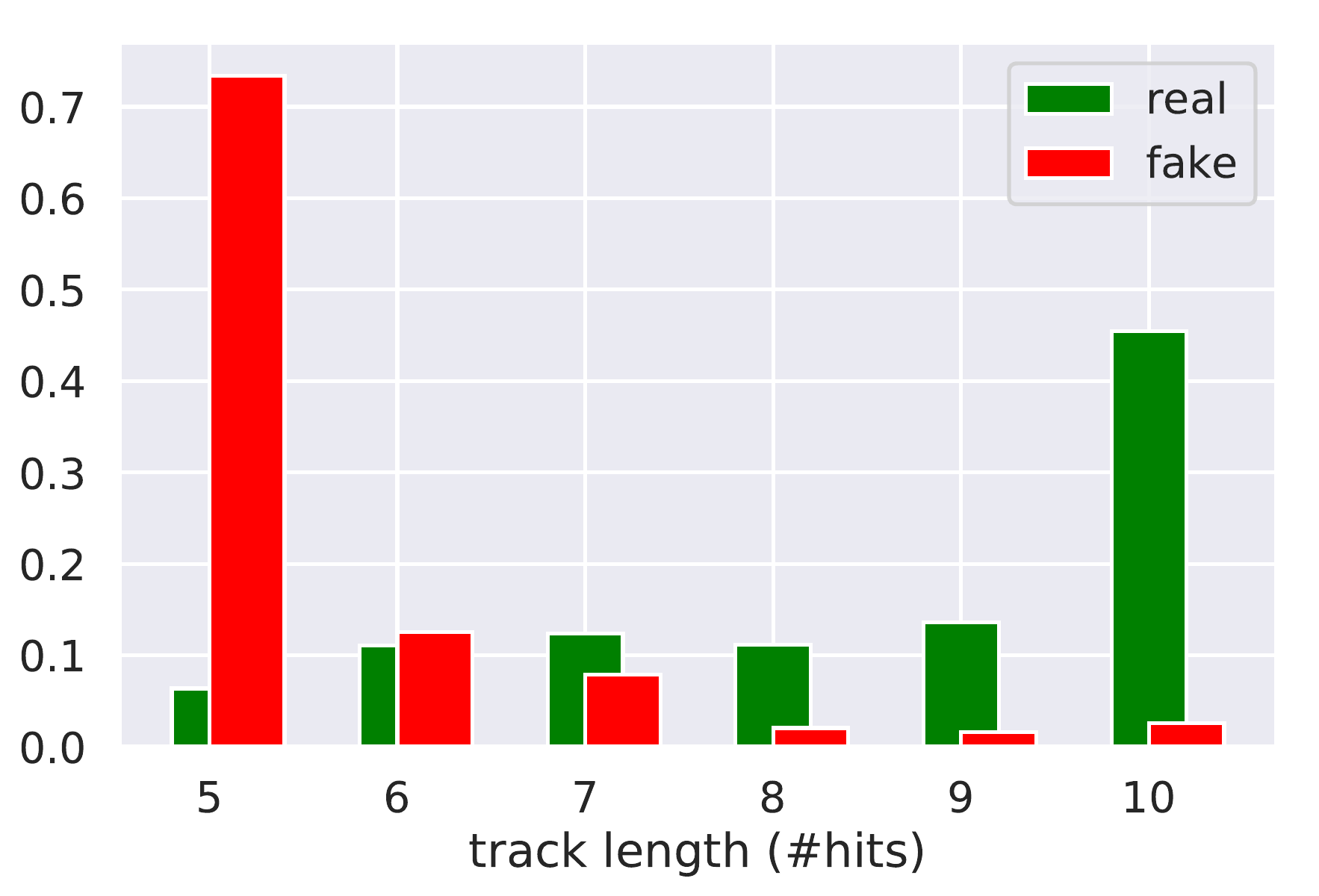}%
\includegraphics[width=.40\textwidth]{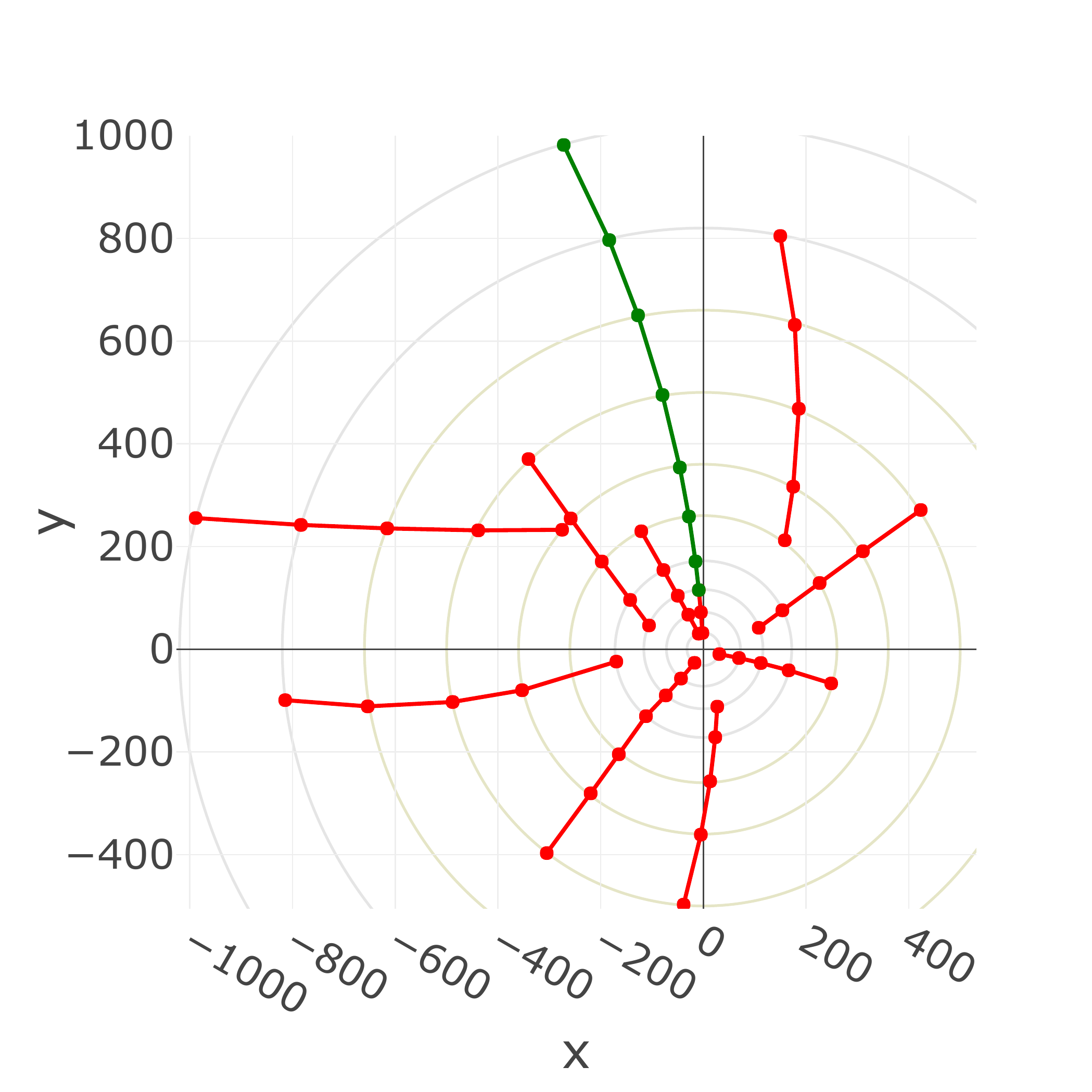}
\caption{The fraction of generated doublets as a function of the number of hits for real (green) and fake (red) tracks 
(left) and an example of a low multiplicity event showing real (green) and fake (red) track candidates (right).}
\label{fig:fake_segments}
\end{figure}

We use purity and efficiency, as defined in Section~\ref{sec:setup-metrics}, to assess the algorithmic performance. Figure~\ref{fig:physics_performance} shows these metrics as a function of the particle multiplicity.
Purity and the TrackML score are well above 90\% across the range, with the efficiency starting close to 100\%, but dropping to about 50\% for the highest occupancies considered.
As the purity drops with increasing occupancy, the number of fake track segments rises (see Figure~\ref{fig:fake_segments}).
The D-Wave machine results are well reproduced by the simulation.

Fig.~\ref{fig:fake_segments} (left) shows the fraction of of real and fake tracks as a function of the number of hits. As the fake tracks tend to have fewer hits, the purity can be improved, with minimal efficiency loss, by requiring barrel tracks to have at least 6 hits. Fig.~\ref{fig:fake_segments} (right) shows examples of fake track candidates.

\begin{figure}[ht!]
\centering
\includegraphics[width=.6\textwidth]{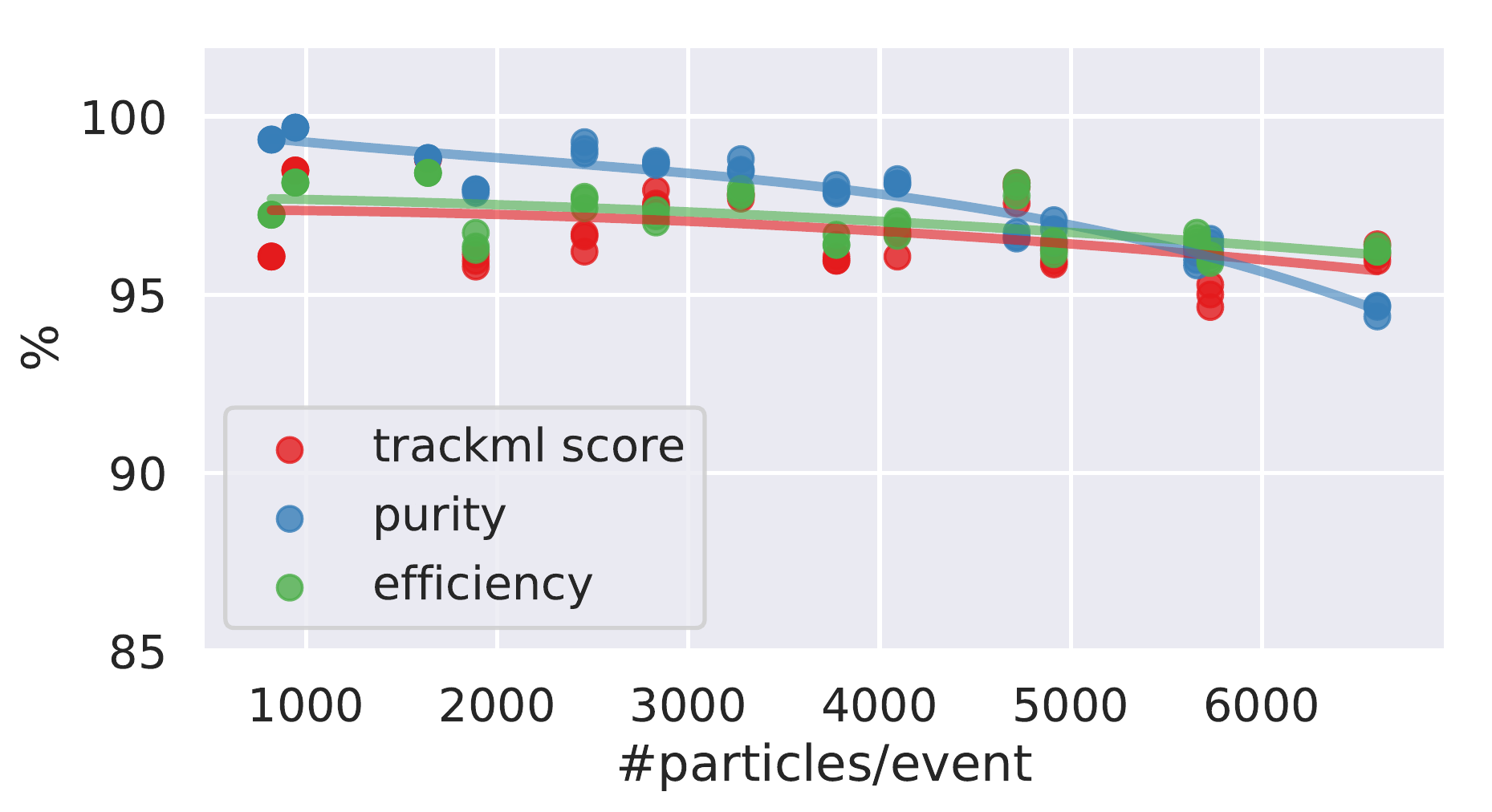}
\caption{The performance of the classical simulator as measured by TrackML score (red), purity (blue), and efficiency (green), as a function of particle multiplicity when including a bias term based on the impact parameter of the triplet.}
\label{fig:impactParmCut}
\end{figure}

The purity can be improved to above 90\% (see Fig.~\ref{fig:impactParmCut}) by adding new properties to the QUBO such as the extrapolated track perigee or impact parameters, but at the cost of biasing the algorithm against tracks with large impact parameters.

We find that the results from the simulator match those of the D-Wave machine rather well. This allows us to use the simulation to tune the parameters for the experiments on the D-Wave machine. No significant impact of noise on the final results is observed.


\subsection{Throughput and Timing}
\label{sec:results-throughput}

Our current experimental setup does not allow to perform detailed timing studies. This is because the devices used are shared, accessed remotely and inherently stochastic.

The generation of the QUBO placement is approximately linear over the range of input doublets considered and takes up to an hour on the largest dataset. However, we expect that the run-time would be improved by code optimization including parallelization. All QUBO solvers scale similarly, with a superlinear running time as a function of occupancy. NEAL is two orders of magnitude faster than qbsolv.

On D-Wave the annealing is run ten times for each sub-QUBO to reduce the impact of noise. There is significant initial setup time on D-Wave, as well as additional overhead due to the time required for minor embedding.


\section{Related Work}
\label{sec:6}
Ref.~\cite{Seddiqi} shows that Quantum Hopfield Associative Memory can be implemented and trained on a D-Wave computer. When training a Hopfield network the optimization goal is to find the set of connection weights that minimizes the network energy for a given set of training patterns. In this work, we used charged particle properties to determine a set of weights and then the set of patterns that minimize the QUBO energy.

Ventura's Quantum Associative Memory (QuAM) is a quantum pattern matching algorithm derived from Grover's search~\cite{Ventura} providing exponential storage capacity~\cite{Shapoval:2019txi}. That algorithm targets pattern recognition algorithms in trigger detectors, while the algorithm discussed here targets offline pattern recognition.

QUBO optimization on D-Wave has been applied to a HEP binary classification problem, a signal/background discriminator for Higgs analyses~\cite{Mott}. While there is no direct relation with the algorithms discussed above, some of the computational challenges are similar.

\section{Discussion}
\label{sec:7}
The main algorithmic innovation reported here is the introduction of a triplet-based QUBO. The richer feature set of a triplet allows the QUBO to achieve greater than 90\% efficiency at track densities which are comparable to HL-LHC\footnote{And two orders of magnitude higher than in Ref.~\cite{stimpfl91}}.
The binary constraints used in the QUBO are based on matching the $p_T$ and $\theta$ track parameters between two triplets. Improvements would likely be achieved by using the full track covariance matrix. Further improvements may come from more refined hyperparameter tuning as well as integration of the detailed geometry and magnetic field description.

When considering throughput, the timing is driven by partitioning the QUBO to fit on the available hardware, given the limited connectivity and the available number of qubits.
The running time of individual sub-QUBOs was observed to be constant. The overall execution time was found to scale with the number of sub-QUBOs. Because of this, we do not currently observe an advantage in running on the D-Wave system. We observed that our large QUBO instances are processed quite efficiently with a particular classical solver. In addition, formulating the problem as a QUBO has the additional advantage to be compatible also with other kinds of special hardware dedicated to the Ising model.



\section{Conclusion}
\label{sec:8}

We have run pattern recognition on events representative of expected conditions at the HL-LHC on a D-Wave quantum computer using qbsolv, and provided a detailed analysis of the algorithm. At low track multiplicity we obtain results with purity and efficiency comparable to current algorithms. We were able to run on events with as many as 6600 tracks. Very good performance was obtained up to approximately 2000 particles per event, after which efficiency remains high, but purity starts to drop. Ideas for future algorithmic improvements were also explored.


%
%

\begin{acknowledgements}
\label{sec:ack}
This research was supported in part by the Office of Science, Office of High Energy Physics, of the US Department of Energy under contract DE-AC02-05CH11231. In particular, support comes from the Quantum Information Science Enabled Discovery (QuantISED) for High Energy Physics program (KA2401032).This research used resources of the National Energy Research Scientific Computing Center (NERSC).
This research also used Ising, Los Alamos National Laboratory's D-Wave quantum annealer.
Ising is supported by NNSA's Advanced Simulation and Computing program.

\end{acknowledgements}

\bibliographystyle{spphys}       
\bibliography{bibliography.bib}   

%
%

\end{document}